\documentclass[twocolumn,prb, showpacs]{revtex4-1}
\usepackage{graphicx}
\usepackage{dcolumn}
\usepackage{bm}
\usepackage{color}

\begin{document}


\title{Surface conduction and $\pi$-bonds in graphene and topological insulator $Bi_2Se_3$}


\author{G. J. Shu$^1$}
\author{F. C. Chou$^{1,2,3}$}
\email{fcchou@ntu.edu.tw}
\affiliation{
$^1$Center for Condensed Matter Sciences, National Taiwan University, Taipei 10617, Taiwan}
\affiliation{
$^2$National Synchrotron Radiation Research Center, Hsinchu 30076, Taiwan}
\affiliation{
$^3$Center for Emerging Material and Advanced Devices, National Taiwan University, Taipei 10617, Taiwan}

\date{\today}

\begin{abstract}
A hybrid orbital model for the topological insulator Bi$_2$Se$_3$ is proposed and compared to that of graphene. The existence of a $\pi$-bond trimer on the Se1 layer at the van der Waals gap and on the surface is thought to be responsible for the unusual surface conduction mechanism found in Bi$_2$Se$_3$ as a topological insulator.  The three $\pi$-bonds are locked as permanent electric dipoles between the Se1 layers in the van der Waals gap as an attractive force for the bulk as a gapped semiconductor.  In contrast, the $\pi$-bond trimers on the surface are unlocked and exhibit two degenerate quantum states, creating a local charge transfer mechanism in the cross-bridge model that is responsible for the surface conduction.  The role of $\pi$-bonds on the Bi$_2$Se$_3$ surface is compared with that in graphene for a similar 2D band structure containing Dirac cones.       
\end{abstract}


\maketitle

\section{introduction}

Topological insulator is a unique state of matter for a band insulator with surface-only conduction and a spin-polarized surface state.  Bi$_2$Se$_3$ is a the second-generation topological insulator with a nearly idealized single Dirac cone at the $\Gamma$-point.\cite{Hasan2010}   The crystal structure of Bi$_2$Se$_3$ (Fig.~\ref{fig:fig01}) can be viewed from the Bi-Se quintuple layers of hcp close packing in space group of R$\bar{3}$m with van der Waals gaps between the Se1-layers.\cite{Huang2012}  The observation of gapless Dirac cone in the surface band structure of topological insulators and in the perfect 2D electron gas of graphene implies that a similar conduction mechanism is expected.\cite{Novoselov2005}  However, the understanding of this new class of material has been based on analogous descriptions of topological invariants, Berry's phase, and the quantum spin Hall effect.  We believe a real space view from the chemical bond perspective should enhance and compliment to the material design significantly.  In particular, we find that there is a great deal of similarity between the graphene and Bi$_2$Se$_3$ surface in terms of the Dirac cone band picture and the existence of $\pi$-bonds. 


\section{Current view of crystal and electronic structures for $Bi_2Se_3$ and graphene }

Graphite has three neighboring carbon atoms ([He]2s$^2$2p$^2$) in each layer, with graphene defined as a monolayer of graphite, which leads to the hybridization of triangular-shaped $sp^2$ orbitals with three coordination sites with the three neighboring carbons in the $\sigma$-bonds and leaves one unpaired electron in the $p$-orbital to form $\pi$-bond with one of the three neighboring carbon atoms statistically.\cite{Neto2009}  Whereas $\pi$-bonds can be formed with any one of the three equivalent neighboring carbon atoms, instead of choosing a particular one to bond with, the bond order is categorized as 1$\frac{1}{3}$ among all carbon atoms on a time-average base.\cite{carbon}  The addition and the dynamic nature of $\pi$-bonds cause graphene to show a stronger bond strength that is manifested by a higher melting point compared with that of diamond; diamond only has one $\sigma$-bond among the carbon atoms in the tetrahedral-shaped $sp^3$ orbitals and can be categorized as having a bond order of 1.  The band picture of graphene has often been described by the uniquely connected Dirac cones between the valence and conduction bands, in which linear dispersion of E(k) suggests the conduction mechanism is describable by the massless Dirac fermions with a relativistic mass of $\sim$ 0.007 $m_e$ and a Fermi velocity of $\sim$ c/300.\cite{Bostwick2006, Katsnelson2006, Geim2009}  The conduction of graphite has often been interpreted through the electron carrier transport in the anti-bonding $\pi^*$-band as the conduction band.\cite{Wallace1947}  Perhaps a more appropriate description of the conduction in graphene would be a chiral tunneling mechanism, which has also been suggested to be the first example of the "Klein paradox"  in condensed matter.\cite{Katsnelson2006}

The interpretation of how can a topological insulator surface conduct electricity relies heavily on the argument of the band inversion phenomenon proposed in heavy elements with strong spin-orbit coupling;\cite{Zhang2009} band inversion for the partially filled $p$ orbitals in Bi and Se may exist under the crystal field field splitting and strong spin-orbit coupling.  Considering the medium electronegativity difference of $\sim$0.5 between Bi and Se, orbital hybridization to form polar covalent bonds among Bi-Se is possible.  The hybrid orbital model for semiconductor silicone has been fully accepted to have four $sp^3$ hybridized $\sigma$-bonds  in tetrahedral coordination among the Si atoms of  atomic electronic configuration of [Ne]3s$^2$3p$^2$, where silicone crystal has a band gap of $\sim$1 eV, just like carbon in the diamond structure.  Since Bi$_2$Se$_3$ has a band gap size of $\sim$0.3 eV in the semiconductor range and both Bi/Se elements are in close packing with octahedral coordination,  we believe that there is no reason not to describe Bi$_2$Se$_3$ with fully hybridized $s$-$p$-$d$ orbitals, especially if the newly proposed hybridization is able to explain the defining physical properties of topological insulators.  In particular, we find the existence of surface $\pi$-bond is unavoidable under the newly proposed $sp^3d^2$ hybridization, and $\pi$-bond is the key component responsible for the electronic structure and physical property interpretation for Bi$_2$Se$_3$ as a topological insulator, which shows great similarity to that of graphene.

\section{crystal structure and chemical bonds of $Bi_2Se_3$}

\begin{figure}
\begin{center}
\includegraphics[width=3.5in]{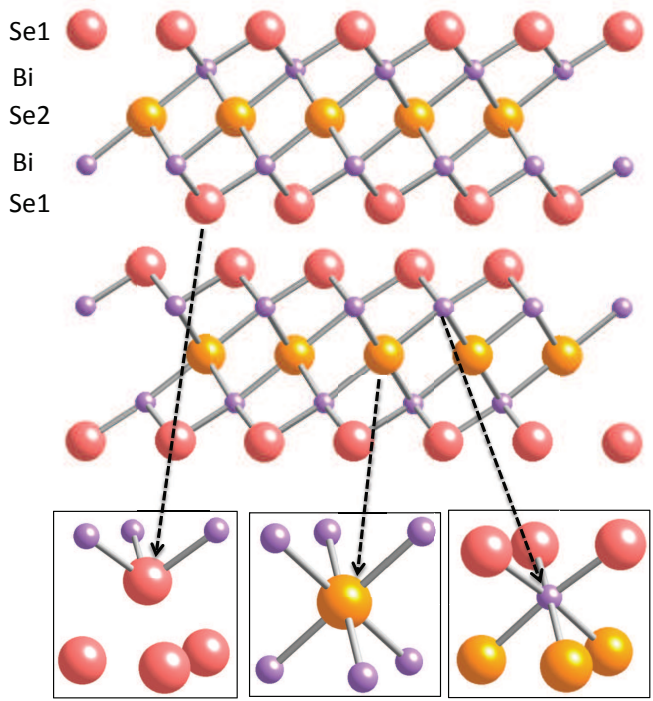}
\end{center}
\caption{\label{fig:fig01}(Color online) The crystal structure of Bi$_2$Se$_3$ is shown with two quintuple layers along the [110] direction for simplicity. The insets show the actual octahedral environment for Se1 at the van der Waals gap, Se2 in the middle, and Bi from left to right, respectively. }
\end{figure}

The crystal structure of Bi$_2$Se$_3$ can be described best using a Bi-Se quintuple as a layer unit, and these quintuple layers are close-packed with symmetry describable with the R$\overline{3}$m space group, i.e., three sliding planes per unit c-axis, as shown in Fig.~\ref{fig:fig01}.   Herein, we propose a real space view of Bi$_2$Se$_3$ through chemical bonds constructed in the hybrid orbital model and use it to explain the defining physical properties observed in Bi$_2$Se$_3$ as a topological insulator.  We will tackle the special physical properties found in Bi$_2$Se$_3$ through real space chemical bonding instead of using the band structure approach.  Parallel interpretations in both real and reciprocal spaces should greatly enhance the understanding of topological insulators and lead to the identification of more materials in this class.

\begin{figure}
\begin{center}
\includegraphics[width=3.5in]{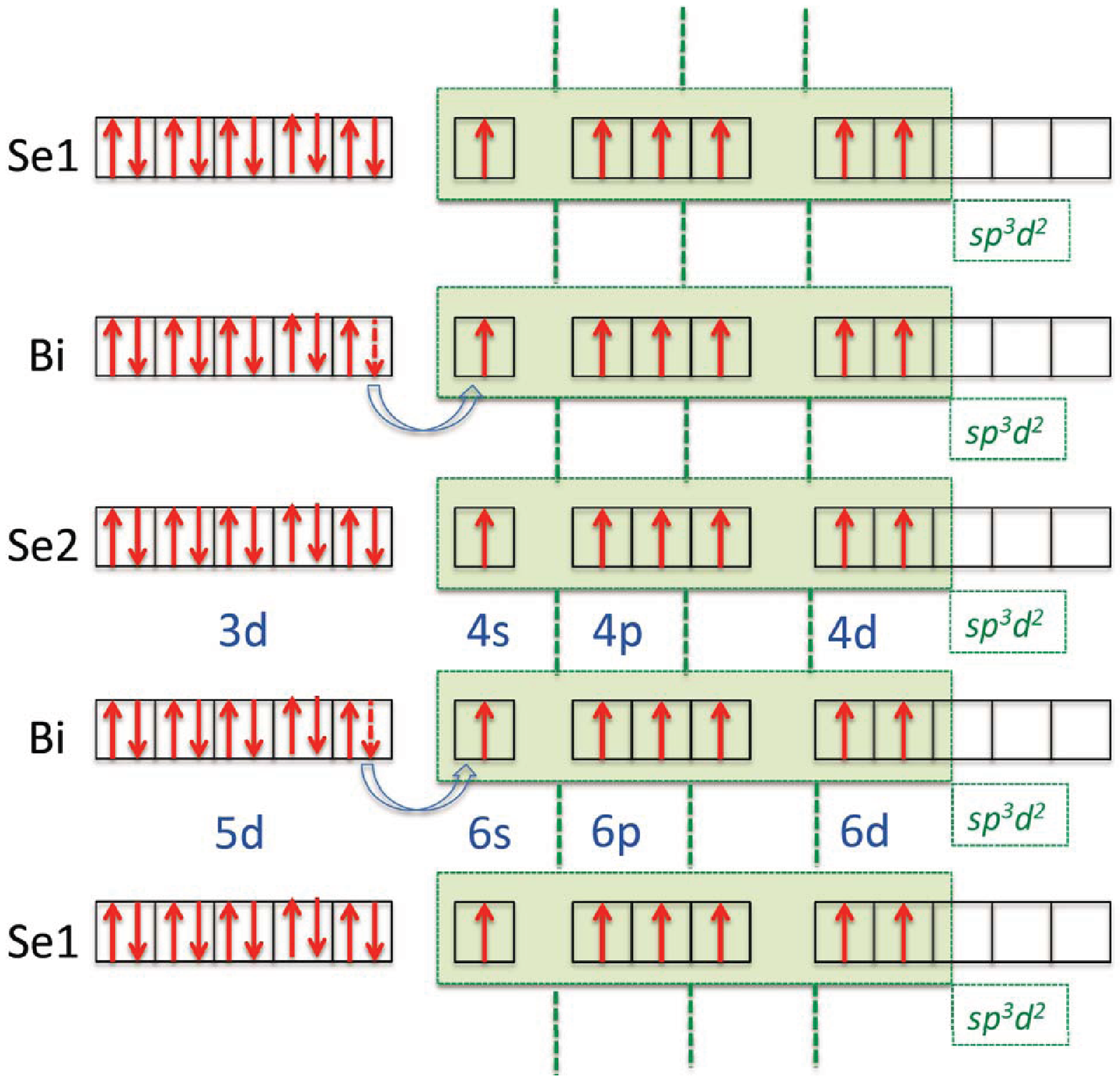}
\end{center}
\caption{\label{fig:fig02}(Color online) The proposed orbital hybridization for one quintuple layer of Bi$_2$Se$_3$. The green dashed lines represent the six hybridized channels of $sp^3d^2$ in the octahedral shape, in which $\sigma$-bonds are formed between Bi-Se, but the three unpaired $\pi$-electrons for Se1 at the van der Waals gap or surface may form a $\pi$-bond trimer with three out of six Se1 atoms in the same layer.} 
\end{figure}

Considering the number of outer-shell electrons in Bi([Xe]4f$^{14}$5d$^{10}$6s$^2$6p$^3$) and Se([Ar]3d$^{10}$4s$^2$4p$^4$) and the close-packed BiSe$_6$-SeBi$_6$ octahedra, the chemical bonding of Bi$_2$Se$_3$ has been proposed based on different choices of orbital hybridization before, such as $p^3d^3$, $sp^3d^2$, or a combination of the two.\cite{Drabble1958, Bhide1971}  Early proposed hybridization models described the unit quintuple layer only without considering the relatively strong van der Waals force.  In addition, crucial information about the Se1 layer surface is missing.  We anticipate that a correct hybrid orbital model for Bi$_2$Se$_3$ should be able to explain the relatively strong van der Waals force, the bulk as a gapped semiconductor, and the unique surface conduction as a topological insulator.  

In each quintuple layer unit of Bi$_2$Se$_3$, there are two different symmetry sites for the Se atoms, one is in the center (Se2), and the other two are at the van der Waals gap (Se1), as shown in Fig.~\ref{fig:fig01}.  Although there are six atoms surround the central one in octahedral coordination for both Bi and Se, the available $s$-$p$-$d$ orbitals for both the Bi and Se atoms should allow $sp^3d^2$ hybridization, as shown in Fig.~\ref{fig:fig02}.  The six outer-shell electrons in Se can easily satisfy the requirement of the six $sp^3d^2$ hybridized orbitals, but we note that the outer-shell of Bi has only five electrons.  One of the $5d^{10}$ electrons should be allowed to move up from the filled $5d$ orbitals to participate in the $sp^3d^2$ hybridization naturally in exchange for the bonding energy gain, especially when the energy difference between the $n=5$ and $n=6$ levels decreases, as expected for heavy elements with a large principal quantum number $n$.  It is clear that both the Bi and Se2 can satisfy the six $\sigma$-bonds required for octahedral shape coordination and create one well-defined quintuple layer.  However, we note that the Se1 at the van der Waals gap will have three electrons remain unpaired.  The three unpaired electrons on Se1 surface should separate as far as possible to 120$^\circ$ apart in a plane to form a trimer to reduce the Coulumb repulsion between them, and three $\pi$-bonds can be formed with three out of the six neighboring Se1 atoms in the same layer.  We call the three $\pi$-bonds in one unit as a $\pi$-bond trimer, and a $\pi$-electron is used to denote the unpaired electron after the $\pi$-bond is broken.

\section{$\pi$-bond conjugated systems}

\begin{figure}
\begin{center}
\includegraphics[width=3.5in]{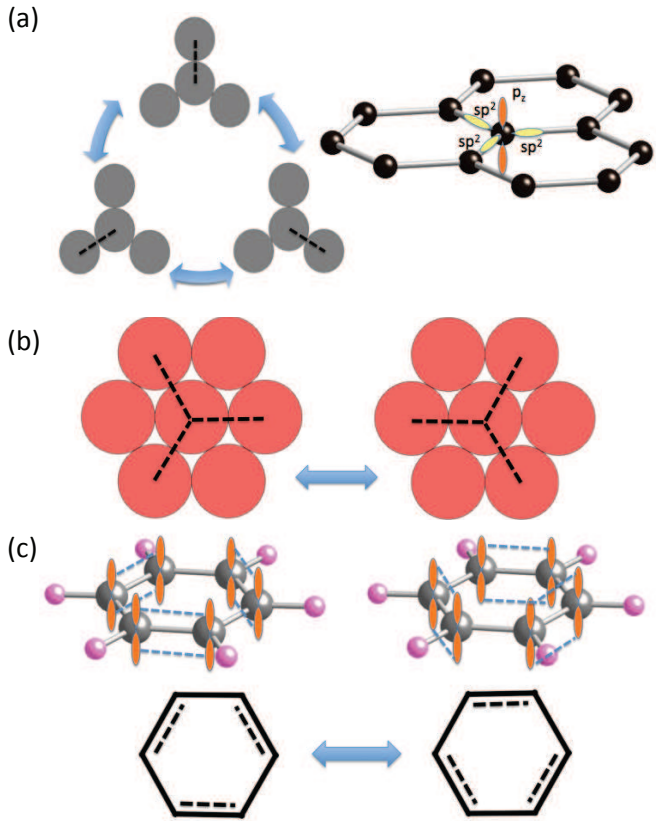}
\end{center}
\caption{\label{fig:fig03}(Color online) The conjugated systems formed with $\pi$-bonds for (a) graphene, (b) Se1 layer on the surface of Bi$_2$Se$_3$, and (c) benzene (C$_6$H$_6$). Graphene has one $\pi$-bond shared by three carbon atoms in $\sigma$-bonds ($sp^2$) as illustrated on the right.  The Se1 layer on the surface of Bi$_2$Se$_3$ has three $\pi$-bonds, forming a trimer with three of the six neighboring Se atoms. There are two degenerated states of $\pi$-bond trimer rotated 60$^\circ$ relative to each other.  Benzene (C$_6$H$_6$) also shows two configurations with three $\pi$-bonds, but this is a closed conjugated system.} 
\end{figure}

For graphene, it is well established in chemistry that statistically, only one unpaired $\pi$-electron in the half-filled $p_z$-orbital would form a $\pi$-bond with one of three neighboring carbon atoms and overlap with one of the three $\sigma$-bonds from the hybridized $sp^2$ orbitals, as shown in Fig.~\ref{fig:fig03}(a).\cite{textbook}  The delocalization of the $\pi$-electron after $\pi$-bond breaking would form a $\pi$-bond with another neighboring carbon instantaneously, and the three $\pi$-bond bonding/nonbonding choices form one conjugated system.  
Although it seems possible for all $\pi$-bonds to orderly lock when considering enthalpy reduction only, this would be less preferable considering the lowering of the total Gibbs free energy for the system that contains competing enthalpy and entropy terms, i.e., the $\pi$-electron configurational entropy gain must also be considered for the system containing easily perturbed, extremely small energy gain/loss through the $\pi$-bond bonding/nonbonding energy exchange.  For Bi$_2$Se$_3$, the Se1 surface can be exposed after cleaving the crystal through the van der Waals gap.  After the Se1 layer is exposed, the three unpaired $\pi$-electrons will form a $\pi$-bond trimer resonating between the two equivalent states as one conjugated system, as illustrated in Fig.~\ref{fig:fig03}(b).     

The entropy factor must be taken into account to determine the comparable impact for a system searching for the true ground state.  In the language of chemical bonding, electrons localized in the valence band are locked in $\sigma$-bonds, and the semi-localized electrons on the Bi$_2$Se$_3$ surface correspond to the much weaker $\pi$-bond.  The $\pi$-bond is constructed through side-to-side minimum orbital overlapping (see Fig.~\ref{fig:fig03}) and is so weak that spontaneous bond breaking is allowed.\cite{textbook}  Another good example is benzene (C$_6$H$_6$), which has three $\pi$-bonds shared by six carbon atoms.  The two equivalent choices of $\pi-$bond configurations and the small $\pi$-bond bonding energy do not preferentially cause the system to settle into either one of the two arrangements, as shown in Fig.~\ref{fig:fig03}(c).\cite{textbook}  Instead, a dynamic conjugated system of raised configurational entropy is the best choice.   

\begin{figure}
\begin{center}
\includegraphics[width=3.5in]{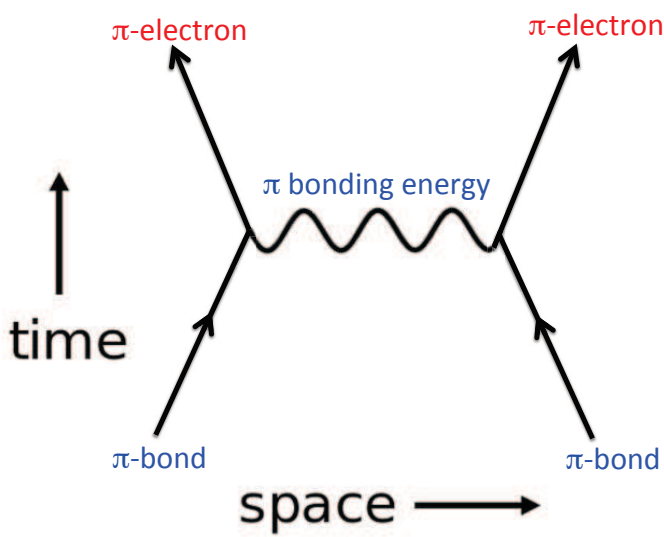}
\end{center}
\caption{\label{fig:fig04}(Color online) An illustration of the time and space relationship for $\pi$-bond, $\pi$-electron, and $\pi$-bond bonding energy exchange as a quasiparticle with a Feynman diagram-like picture. }
\end{figure}

We might describe such a dynamic $\pi$-bond energy exchange in time and space using a Feynman diagram-like picture, as illustrated in Fig.~\ref{fig:fig04}.  Similar entropy-enthalpy competition has been applied to the uncertain $\pi$-bond position among the three choices of the directional $\sigma$-bonds, as shown in Fig.~\ref{fig:fig03}(a) for graphene, i.e., although it takes energy to break the existing $\pi$-bond at a specific time and position, the new $\pi$-bond formation at a different position with a configurational entropy gain would still be able to lower the total Gibbs free energy for the remaining terms within $\Delta G=\Delta \textbf{H}-T\Delta S$.  In the language of thermodynamics, which addresses the initial and final states only, $\Delta \textbf{H}$=0 always, although the $\pi$-bond bonding energy is nonzero and exchanged constantly.  In nature, the dynamic conjugated system is preferred over the static choices (whether ordered or not) for systems with energy gaps as small as the weak $\pi-$bond bonding energy.

\section{van der Waals gap in $Bi_2Se_3$ and bilayer graphene}

\begin{figure}
\begin{center}
\includegraphics[width=3.5in]{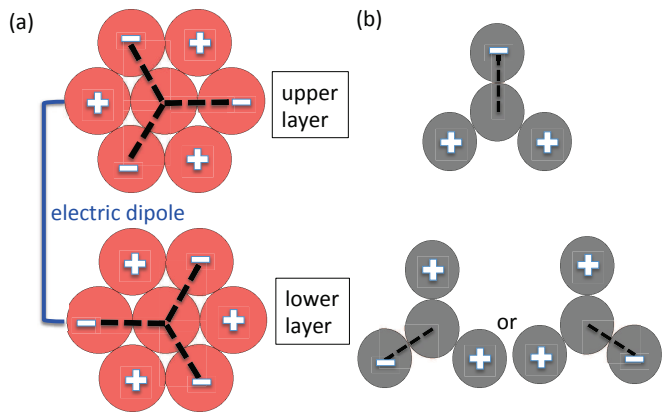}
\end{center}
\caption{\label{fig:fig05}(Color online) The electric dipoles formed with $\pi$-bonds in the van der Waals gap between (a) two Se1 layers in the van der Waals gap of Bi$_2$Se$_3$, and (b) between the bilayer graphene.  All six electric dipoles are locked definitely for the former, whereas one of the three possible dipoles must be frustrated in bilayer graphene to weaken its attractive force.  Note that these two layers are actually shifted for hexagonal close packing in reality.}
\end{figure}

Although the quintuple layers are stacked with sliding 2D triangular lattices to form close packing, as shown in Fig.~\ref{fig:fig01}, the bond length of Se1-Bi (2.85$\AA$) in the virtual octahedron of (Se1)Bi$_3$(Se1)$_3$ is significantly different to that of Se2-Bi (3.08$\AA$) in the (Se2)Bi$_6$ octahedron.\cite{Nakajima1963}  In addition, the Se1 atoms are arranged in a 2D triangular lattice, i.e., there are six Se1 atoms surrounding each Se1, as shown on the opposite sides of the van der Waals gap shown in Fig.~\ref{fig:fig05}.  For each Se1 atom exposed to the van der Waals gap, there are three unpaired electrons waiting to bond.  These electrons are called $\pi$-electrons before a $\pi$-bond  is formed with the neighboring Se1. The three half-filled $sp^3d^2$-orbitals can form three $\pi$-bonds with three of the six neighboring Se1 atoms on the surface, with a weaker bonding energy due to the minimum orbital overlapping compared with the $\sigma$-bond.  Considering the Coulomb repulsion among the three unpaired electrons, a symmetric triangular distribution should be preferred with two available degenerate states, which is defined as a $\pi$-bond trimer for simplicity.  When two Se1 layers approach each other in the van der Waals gap, a total of six electric dipoles can form in such a way that two $\pi$-bond trimers are locked  in a rotation that is 60$^\circ$ relative to the other, which could be viewed as the source of an attractive force in the van der Waals gap.   

It is interesting to compare the strength of the van der Waals force between the bilayer graphene and Bi$_2$Se$_3$ through $\pi$-bond and electric dipole formation.  Considering that the surface of Bi$_2$Se$_3$ is composed of completely free $\pi$-bond trimers, each $\pi$-bond trimer has two possible choices in the next move through clockwise or counterclockwise motion.  When a second identical Se surface approaches just as in van der Waals gap formation, the second Se surface has no choice but to form six permanent electric dipoles to avoid Coulomb repulsion and locks to the first Se surface accordingly, as shown in Fig.~\ref{fig:fig05}(a). In contrast, the $\pi$-bond in each graphene layer has three choices at any instance. The approaching second graphene layer would always have one of the three possible dipoles frustrated relative to that of  the first layer, which leads to a weaker attractive force.  The weaker van der Waals force is the key property that makes the mechanical extraction of a monolayer of graphite (graphene)  possible,\cite{Novoselov2005} whereas we find that the same mechanical cleavage method failed to prepare a mono quintuple layer from single crystal Bi$_2$Se$_3$.  Considering the quintuple layer thickness of $\sim$7 $\AA$ and the van der Waals gap size of $\sim$2.5 $\AA$, it is not surprising to find that for Bi$_2$Se$_3$, a gapless Dirac cone can be observed and defined as a topological insulator for only up to five layers of quintuples,\cite{He2010} which must be closely related to the inter $\pi$-bond trimer distance across the actually exposed Se1 surface, i.e., whether $\pi$-bond trimer is locked is closely related to the gap opening at the Dirac cone for surface conduction.  We will describe this concept in greater detail in the following section.

\section{$\pi$-bond conduction mechanism}

\begin{figure}
\begin{center}
\includegraphics[width=3.5in]{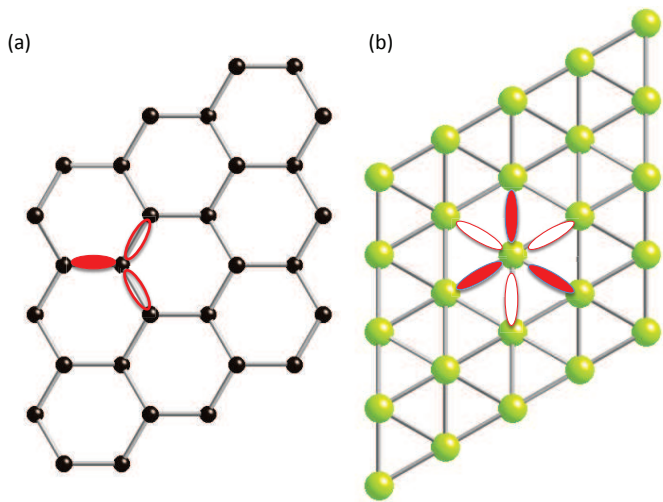}
\end{center}
\caption{\label{fig:fig06}(Color online) A cross-bridge model can be used to explain the 2D conduction mechanism for a snapshot of (a) graphene, in which an ($1\sigma$+$\frac{1}{3}$$\pi$)-bond is found among all carbons on time-average basis, and (b) a Se1 surface of Bi$_2$Se$_3$, in which the solid and empty lobes represent the two choices of $\pi$-bond trimers that lead to $\frac{1}{2}$$\pi$-bonds on a time-average basis among all Se1 atoms.  An infinite percolation route can be constructed for three-way connection using one constantly rotating $\pi$-bond bridge for graphene, and for a six-way connection with one rotating $\pi$-bond trimer for Se1 layers on the surface of Bi$_2$Se$_3$. }
\end{figure}

We find it is necessary to re-examine the conduction mechanism of graphene and Bi$_2$Se$_3$ surfaces using the unique quantum nature of $\pi$-bonds in real space, instead of following the conventional view of the $\pi^*$ conduction band picture that treats unpaired $p_z$ electrons as itinerant carriers.\cite{Wallace1947}  Unlike the conventional view that distinguishes electrons in either the localized state in the valence band or in the itinerant state in the conduction band, the $\pi$-electron has a unique dual role of coexisting localized (spin) and itinerant (charge) characteristics in the time window, as expected for the chiral electrons described by the quantum spin Hall effect, and the surface bandgap is so small that quantum tunneling becomes possible,\cite{Katsnelson2006} which is also describable by the impact of entropy on the system.  The constant $\pi$-bond bonding/nonbonding action among the three choices in space for each atom site locally can be illustrated with a Feynman diagram-like picture in space and time as shown in Fig.~\ref{fig:fig04}, where $\pi$-bond bonding energy can be exchanged to form a quasiparticle-like conjugated system.  

The Dirac cone with a linear dispersion of E(k) also implies that the surface 2D conduction mechanism is better described by Dirac fermions following relativistic dynamics with zero effective mass.\cite{Katsnelson2006}  Such massless transport implies that itinerant electrons are not the actual carriers for electricity transport; instead, the spontaneous and constant energy exchange should be responsible for the conduction of electricity.  To construct a continuous route for the electricity flow based on the local $\pi$-bond constant bonding/nonbonding action, as described in Fig.~\ref{fig:fig04}, 
we propose a cross-bridge model to explain the unique conduction mechanism for both graphene and Bi$_2$Se$_3$ surfaces in real space.  

In the case of graphene, only one $\pi$-bond can form  with one of the three neighboring carbon atoms at any instance, but a continuous percolation route can always be formed at any instance, as shown in Fig.~\ref{fig:fig06}(a).  In the case of the Se1 layer on a Bi$_2$Se$_3$ surface, there is one $\pi$-bond trimer (three $\pi$-bonds) shared with the six neighboring Se atoms, and a continuous percolation route can also always form, as shown in Fig.~\ref{fig:fig06}(b).  Graphene conduction can be described with one constantly shifting $\pi$-bond bridge among the three $\sigma$-bonds, which seems to have an (1$\sigma$+$\frac{1}{3}$$\pi$)-bond between any two carbons on a time-average basis, but the action of $\pi$-bond shifting is actually discrete in time and localized on the same atom.  Similarly, the bond order among Se1 atoms on the surface of Bi$_2$Se$_3$ can also be viewed as a $\frac{1}{2}$$\pi$-bond on a time-average basis, although the $\pi$-bond trimer is constantly shifting between the two choices of clockwise and counterclockwise rotation discretely in time, centered on each Se1 atom.  

It is also interesting to note that such a cross-bridge model has been applied in the field of biomechanics to explain the  continuous sliding function of muscles.  The cross-bridge model explains how a myosin filament with only three myosin heads (separated 120$^\circ$) can continuously attach the actin filaments arranged in a close-packed hexagonal 2D lattice in a cross-sectional view.\cite{Redaelli2001}  It must be noted that the existence of $\pi$-bonds does not guarantee electric conduction.  For example, benzene (C$_6$H$_6$) has three $\pi$-bonds within one loop, as shown in Fig.~\ref{fig:fig03}(c).  The two choices of $\pi$-bond configurations form one dynamic conjugated system of two equivalent states, but the open and close actions of $\pi$-bonds do not create an open and continuous percolation route for charge transfer.\cite{textbook}  In contrast, conduction becomes possible in a conducting polymer that contains both a $\pi$-bond conjugated system and an open and  continuous percolation route.\cite{Bryce1991}

How do graphene and the Bi$_2$Se$_3$ surface conduct electricity in the absence of the conventionally defined itinerant electron carriers?  First, there is a percolation network of nonbonding $\pi$-electrons that dynamically exists all the time according to the cross-bridge model discussed above.  In other words, the $\pi$-bond trimers must be unlocked from the electric dipole formed in the van der Waals gap first.  Second, although the $\pi$-electron remains in the local vicinity of each atom, the energy is allowed to be transferred between these atoms through a relay-type energy exchange, i.e, although it takes $\pi$-bond bonding energy to excite the electron to the nearly-free $\pi$-electron state before it bonds again statistically, the same amount of $\pi$-bonding energy is exchanged between atoms like a quasiparticle, as illustrated in Fig.~\ref{fig:fig04}.  This type of electricity transport does not require actual itinerant electron carrier movement, only local $\pi$-bond bonding energy exchange.  Third, analogous to the transverse wave in water, the locality of $\pi$-electrons can be viewed in parallel to the hydrogen-bond coupling among water molecules.  Unless there exists a $\pi$-bond that is local and with weak bonding energy, such an energy relay mechanism would not be possible.  The bonding/nonbonding mechanism of the $\pi$-bond loses no energy in the energy relay process when considering the initial and final states only, i.e., $\Delta \textbf{H}=0$ thermodynamically, which is in agreement with the quantum tunneling description for the massless Dirac fermions.\cite{Katsnelson2006}  

\section{Conclusions}

In summary, we have constructed a hybrid orbital model for Bi$_2$Se$_3$ based on crystal symmetry, atomic coordination, and the outer-shell electrons.  The $\pi$-bond trimer for Se on each quintuple layer unit can be used to successfully explain the gapped semiconducting behavior for the bulk and the conducting surface.  We believe this real space view from a chemical bond perspective should be complimentary to the understanding of topological insulator that has often been approached previously using the quantum spin Hall effect and band topology. The chemical bond approach could help us design additional topological insulators more intuitively and shift our focus to materials containing an open conjugated system that is constructed with $\pi$-bonds on the surface.  It is clear that more experimental works are necessary to trace phenomena related to the existence and behaviors of $\pi$-bonds, especially more theoretical works to bridge the real and reciprocal views of topological insulators in parallel to graphene. 

\section*{Acknowledgment}
FCC acknowledges support from NSC-Taiwan under project number  NSC 101-2119-M-002-007.  GJS acknowledges support from NSC-Taiwan under project number NSC 100-2112-M-002-001-MY3.




\end{document}